# Orthorhombic *Fddd* Network in Diblock Copolymer Melts

Mikihito Takenaka*, Tsutomu Wakada, Satoshi Akasaka, Shotaro Nishitsuji, Kenji Saijo, Hirofumi Shimizu, and Hirokazu Hasegawa

Department of Polymer Chemistry, Graduate School of Engineering, Kyoto University, Kyoto, 615-8510, JAPAN

Soft-materials such as block copolymers, surfactant and liquid crystals exhibit variety of ordered microstructures. Among them, the phase diagrams of diblock copolymers have been extensively investigated both experimentally and theoretically. Matsen and Shick[1] calculated the phase diagram of diblock copolymer melts by using self-consistent field theory (SCFT) and predicted that the phase diagram contains four types of structures: sphere packed in body-center-cubic, hexagonally-packed cylinders, lamellar and double-gyroid network. Khandpur et al. investigated the phase diagram of poly(styrene-b-isoprene) diblock copolymer in detail and show the good agreement between experimental results and the phase diagram calculated by SCFT[2]. Then, in order to seek more variety of ordered structures, several researches on the phase behaviors of triblock copolymers have been done[3-10]. Among these studies a remarkable finding has been done by Bailey et al.[3]. They identified a noncubic network morphology, the *Fddd* ($O^{70}$) structure in poly(isoprene-*b*-styrene-*b*-ethyleneoxide) (ISO) triblock copolymers. Epps et al. confirmed that the $O^{70}$ phase is an equilibrium phase in triblock copolymers[6, 7]. Tyler et al. studied the phase behavior of triblock copolymers as a function of composition at given conditions, where the Flory-Huggins interaction parameters, polymerization index, and statistical segment length are fixed, by using SCFT[11]. The calculated phase diagram is quite similar to that of ISO and the $O^{70}$ phase is found to be located between gyroid, lamellae, and alternating gyroid phases as an equilibrium phase. Tyler et al. also found the $O^{70}$ phase between gyroid, lamellae, and cylinder phases with SCFT calculation although the region of the $O^{70}$ phase is limited within quite a narrow temperature region[11]. Since the SCFT calculation has successfully predicted the phase diagram of block copolymers melts, we can anticipate the existence of the $O^{70}$ phase in the phase diagram of diblock copolymer melts. We, thus, studied the phase behavior of poly(styrene-*b*-isoprene) by using small-angle X-ray scattering (SAXS) and transmission electron microscope (TEM) in detail. The SCFT calculation predicted that diblock copolymer melts with slight asymmetric volume fraction $f$ ( $0.42 < f < 0.5$) exhibits the $O^{70}$ phase and Khandpur et al. found that the isoprene rich region shows the rich variety of ordered structure such as gyroid, HPL and so on. We thus studied an isoprene-rich S-I diblock copolymer. The S-I used here has volume fraction of isoprene $f_{PI}=0.63_8$, the number –averaged molecular weight $M_w=2.64\times10^4$ and the heterogeneity index $M_w/M_n= 1.02$. The S-I was synthesized by an anionic polymerization method. SAXS experiments were conducted on BL-15A at KEK, Japan. The wavelength used here is 1.54Å and the distance from sample to detector is 2000mm. We used imaging plates as a detector. We observed the morphology of the S-I diblock copolymer with a transmission electron microscope (JEM-2000EX, JEOL Co., Ltd) at an acceleration voltage of 160kV.

    Figure 1 shows the temperature dependence of azimuthally-averaged SAXS profiles of the S-I. At 230˚C, a broad peak is observed which corresponds to disordered phase. Below 170˚C, several distinct peaks appear in the profiles, indicating that the S-I is in its ordered state. Except for T=160˚C, and T=135˚C, The peak positions at the other temperatures coincide each other. In the case of T=150˚C,

the SAXS peaks are located at $q/q_m$ = 1, 1.22, 1.56, 1.72, 1.81, 1.94, 2, 2.18, 2.29, 2.49, 2.75, 2.93, 3. This series of peaks suggests that the lattice structure of the S-I is an orthorhombic. The peak positions of orthorhombic lattices are calculated by

$$q_{hkl} = 2\pi[h^2/a^2 + k^2/b^2 + l^2/c^2]^{1/2}, \qquad (1)$$

where $a$, $b$, and $c$ are unit cell parameters and $h$, $k$, and $l$ are Miller indices for $a$, $b$, and $c$, respectively. We estimated $(a:b:c) = (1:2.02:3.47)$ and the Miller indices indicated by arrows in Fig.1. Similar to the case of $O^{70}$ structures of ISO triblock copolymer melts by Epps et al.[6], the 113, 131, 040, 133, 202, 220, 222, 044, and 135 positions agree with the peak positions in the SAXS profile. Moreover, 242, 313, 315, and 333 positions which are allowed for $O^{70}$ structures are found as peaks. These agreements suggest that the S-I diblock copolymer forms $O^{70}$ structures. In the case of ISO triblock copolymer melts, 1st order peak becomes relatively broader since the 111, 022, and 004 reflections slightly differ each other. On the other hand, the 1st order peak is relatively sharp. This difference is attributed to the difference in unit cell parameters ratios. The unit cell parameters ratio of the ISO triblock copolymer is typically $(a:b:c) = (1:1.98:3.40)$ and the calculated peak ratios $q_{022}/q_{111}$ and $q_{004}/q_{111}$ with Eq.(1) for the ISO triblock copolymer melts are, respectively, 1.01 and 1.02. On the other hand, $q_{022}/q_{111}$ and $q_{004}/q_{111}$ with Eq.(1) for the S-I diblock copolymer melt are, respectively, 0.99 and 1.00 so that the 004 position of the S-I diblock copolymer melt is closer to 111 position than those of the ISO triblock copolymer melts. Thus, the 1st order peak is relatively sharp. It should be noted that the unit cell parameters ratio of the S-I diblock copolymer melt agrees with $(1:2:2\sqrt{3})$ obtained in SCFT calculation by Tyler et al.[11]. Tyler et al. suggested that this ratio causes the near-coincidence of 004, 111, and 022 positions, which contributes to the stability of $O^{70}$ structures.

Figure 2 shows the representative TEM micrograph of the S-I diblock copolymer annealed at 170˚C. In the micrograph, the dark part corresponds to isoprene domains since the sample was stained with $O_sO_4$. In the micrograph, staggered rows of bright ovals are interconnected with trivalent junctions and the observed patterns does not hold 3-fold and 4-fold symmetry. These features agree with those obtained for $O^{70}$ structures of ISO triblock copolymer melts[6]. Thus, TEM observation also confirmed that the obtained structure correspond to $O^{70}$. Although we do not show here, birefringence appears below 210˚C. This result also supports the existence of $O^{70}$.

At 137.5˚C, the peak positions ratio of the SAXS profile are 1, 2, and 3 suggesting that the S-I diblock copolymer melt forms lamellar structurse at this temperature. It should be noted that in addition to the peaks from lamellar structures, the weak peaks appear at $q/q_m$ = 1.23 and 1.95 which coincide to the peak positions observed in $O^{70}$ structures. This fact indicates that the $O^{70}$ structure coexists with lamellar structures at this temperature although the lamellar structures are dominant and stable at this temperature.

A strange feature appears in SAXS profiles at 165˚C and 160˚C. The second peak position splits into two peaks at $q/q_m$ = 1.22 and 1.14, while the higher order peaks agrees with the peaks observed at 150˚C, 170˚C and 172˚C. Since gyroid structures has a second peak at $q/q_m$ = 1.15, the coexistence of gyrroid and $O^{70}$ structures is one of the possible reason of these peaks. However, we can not observe the higher order peaks originating from gyroid structures. Thus we can not identify definitively the structures present in this temperature region. This point will be addressed in future work.

Finally, we plotted the points of $O^{70}$ structures in the phase diagram obtained by Khandpur et al[2]. We used the following temperature dependence of $\chi$ parameter reported in their paper:

$$\chi = 71.4/T - 0.0857. \qquad (2)$$

The $O^{70}$ region is located between Lamellar, Gyroid and HPL regions. This location of the $O^{70}$ region agrees with the SCFT phase diagram obtained by Tyler et al[11]. According to the SCFT phase diagram, we anticipate that the S-I with $f$ being slightly larger than 0.64 exhibits gyroid- $O^{70}$-lamellar transition with decreasing temperature. We will investigate the phase diagram around $f$=0.64 in detail and confirm the regions of the $O^{70}$ structures and the order-order transition between the morphologies.

**Figure captions**

Figure 1  Azimuthally-averaged SAXS profiles for S-I at various temperatures. Samples were first annealed at 230˚C and then annealed at each temperature prior to measurements. Indices of SAXS profile for 170˚C are calculated for *Fddd* structures with unit lattice cell constant. The SAXS profiles are shifted vertically for clarity.

Figure 2  TEM image of the S-I. Sample was subjected to the anneal at 230˚C followed by annealing at 170˚C and then quenched from 170˚C into liquid nitrogen. Scale bar is 10nm. Network structure consisting of staggered rows of bright ovals are interconnected with trivalent junctions.

Figure 3 Phase diagram of S-I diblock copolymers. Open square, filled triangle, and cross symbols correspond to disordered state, $O^{70}$ structure and lamellar structures, respectively. Open triangle, corresponds to the coexistence of gyroid and $O^{70}$ structures. Adapted from Khandpur et al.[2].

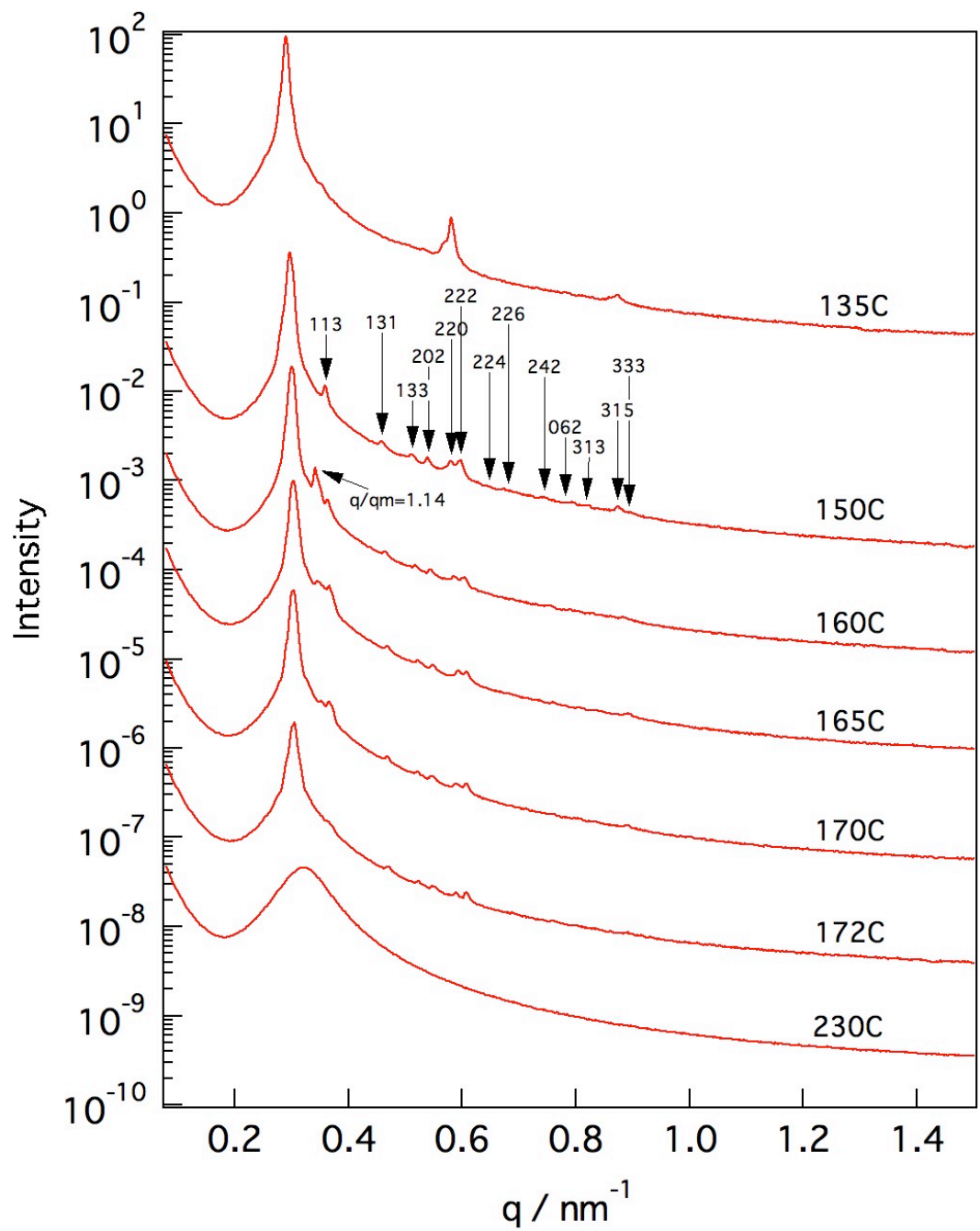

Figure 1

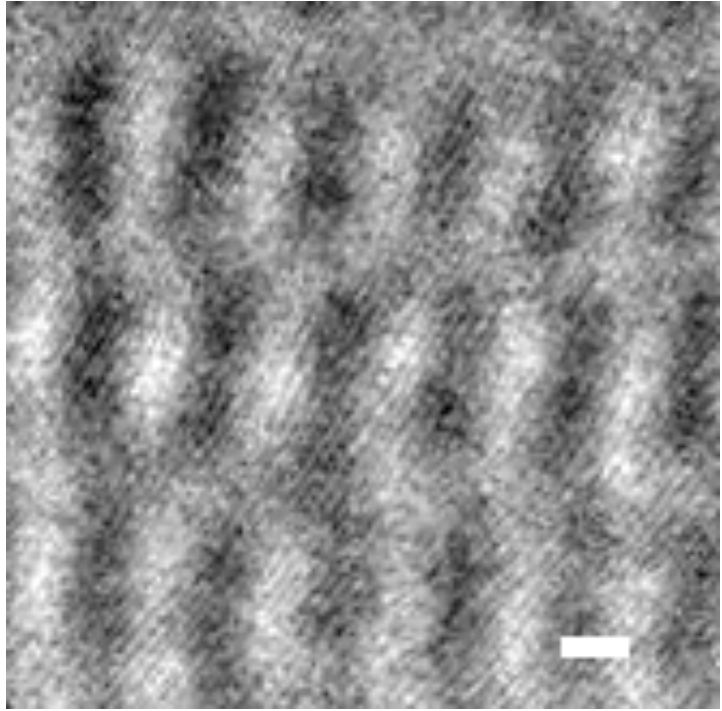

Figure 2

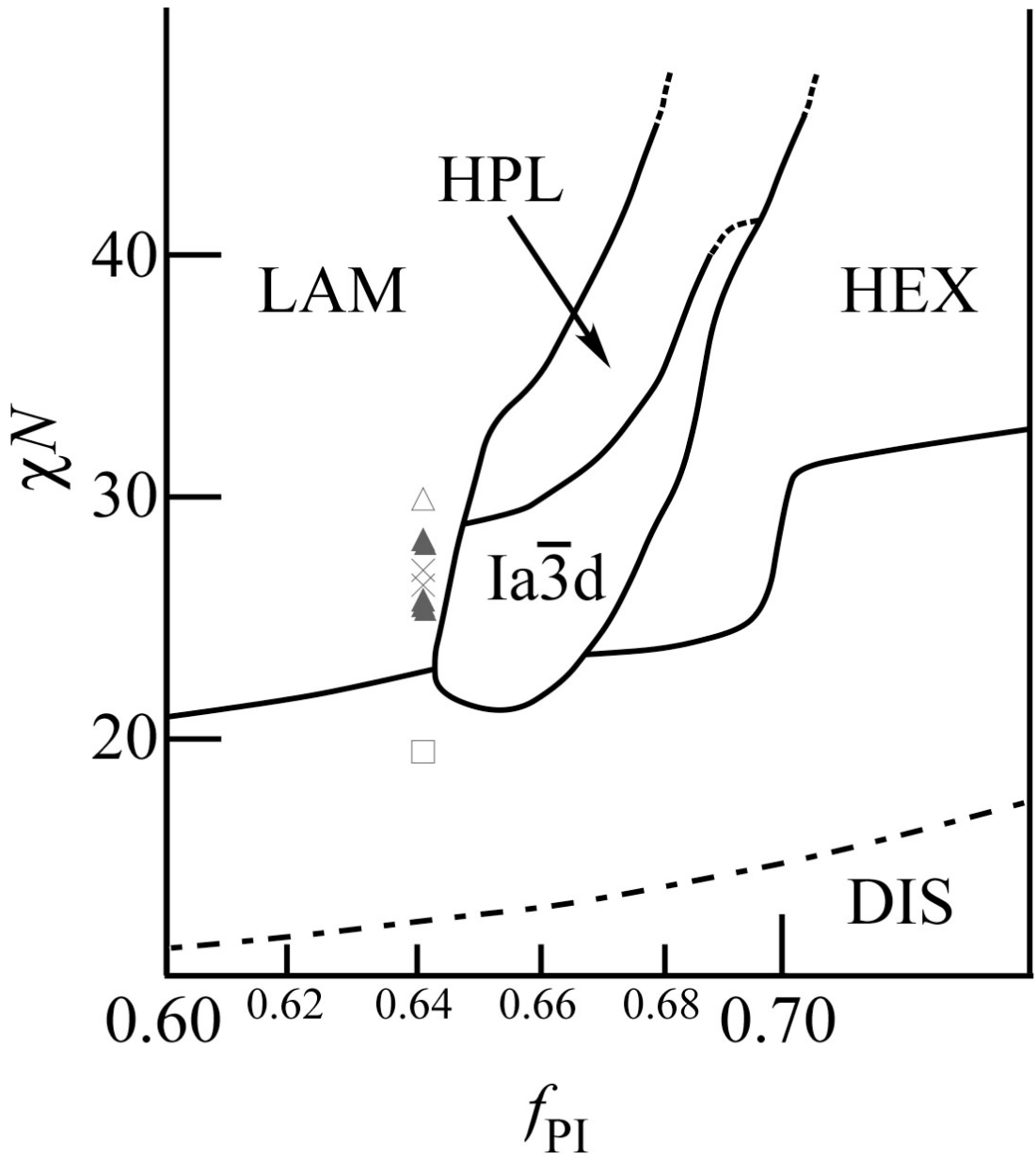

Figure 3